\documentclass[letter]{jpsj2} 
\title{Observation of Half-Integer Quantum Hall Effect in Single-Layer Graphene\\ Using Pulse Magnet}

\author{
	Satoru \textsc{Masubuchi}$^{1}$\thanks{E-mail: msatoru@iis.u-tokyo.ac.jp}, 
	Ken-ichi \textsc{Suga}$^{2}$,
	Masashi \textsc{Ono}$^{1}$,\\
	Koichi \textsc{Kindo}$^{2}$,
	Shojiro \textsc{Takeyama}$^{2}$,
	and Tomoki \textsc{Machida}$^{1,3}$\thanks{E-mail: tmachida@iis.u-tokyo.ac.jp}
}

\inst{
$^{1}$Institute of Industrial Science, University of Tokyo, 4-6-1 Komaba, Meguro-ku Tokyo 153-8505

$^{2}$ Institute for Solid State Physics, University of Tokyo, 5-1-5 Kashiwanoha, Chiba 277-8581

$^{3}$ Institute for Nano Quantum Information Electronics, University of Tokyo, 4-6-1 Komaba, Meguro-ku Tokyo 153-8505}

\abst{
We report on the magnetotransport measurement on a single-layer graphene in pulsed magnetic fields up to $B$ = 53 T. With either electron- or hole-type charge carriers, the Hall resistance $R_\textrm{H}$ is quantized into $R_\textrm{H}$ = $(h/e^2)\nu ^{-1}$ with $\nu$ = $\pm$2, $\pm$6, and $\pm$10, which demonstrates the observation of a half-integer quantum Hall effect (QHE). At $B$ = 50 T, the half-integer QHE is even observed at room temperature in spite of a conventional carrier mobility $\mu$ = 4000 cm$^2$/Vs.}

\kword{graphene, quantum Hall effect, pulsed magnetic field technique}

\begin{document}
\maketitle

\newpage
Graphene, a single atomic layer of graphite, has attracted significant attention after the experimental achievement of depositing single-layer graphene on silicon wafers \cite{Novoselov04}. Since then, the remarkable transport phenomena of massless two-dimensional Dirac fermions have been reported: the half-integer quantum Hall effect (QHE) \cite{Novoselov05, Zhang05, Peres06}, the strong suppression of weak localization \cite{Morozov06}, and the room-temperature QHE \cite{Novoselov07}. These unique electronic properties of graphene are derived from its unique band structure, that is, the linear conduction and valence bands touch at Dirac neutrality points \cite{Wallace47, McClure56}. When a strong magnetic field $B$ is applied perpendicularly to the graphene layer, its electronic spectrum is quantized into Landau levels (LLs). Since the $N$ = 0 LL at the Dirac point is shared by holes and electrons, the Hall resistance $R_\textrm{H}$ shows a half-integer shifted sequence of quantization, such that  $R_\textrm{H}$  = $(4h/e^2)(N + 1/2)^{-1}$ where $N$ is an integer.

Recent theoretical studies of the half-integer QHE in graphene have focused on the valley and spin splittings in LLs due to many-body effects of massless Dirac fermions. Valley-splitting QH states at the LL filling factors $\nu$ = $\pm$1 \cite{Nomura06} in the $N$ = 0 LL were anticipated at strong magnetic fields $B$ $\sim$ 50 T \cite{Note01}, and experimental studies under DC high magnetic fields revealed $\nu$ = $\pm$1 QH states \cite{Zhang06, Jiang07}. The existence of valley-splitting $\nu$ = $\pm$3 and $\pm$5 QH states in the $N$ = $\pm$1 Landau levels due to many-body effects at magnetic fields $B$ $\sim$ 300 T \cite{Note01} is under considerable debate \cite{Nomura06, Gusynin06}. Also, the fractional QHE of composite Dirac fermions is expected at higher magnetic fields \cite{Toke06, Apalkov06}. Therefore, experimental studies of the half-integer QHE in graphene at high magnetic fields have great importance for the study of the fundamental electronic properties of graphene. Such a strong magnetic field can be attained only by pulsed magnets, whereas the present-day DC magnetic field is limited to 45 T. Up to now, the measurements of the half-integer QHE in pulsed magnetic fields have been reported only for the two-terminal magnetoresistance $R_\textrm{2p}$ \cite{Krstic08}. The observation of the correctly quantized half-integer QHE in pulsed high magnetic fields has not been reported yet.

In this work, we report on the magnetotransport measurement on a single-layer graphene multiterminal device in pulsed magnetic fields up to $B$ = 53 T. In either hole- or electron-doped graphene, the Hall resistance is quantized into  $R_\textrm{H}$ = $(h/e^2)\nu ^{-1}$ with integer values $\nu$ = $\pm$2, $\pm$6, and $\pm$10, indicating the observation of half-integer QHE. This result is contrast to the measurement of the two-terminal magnetoresistance of graphene in pulsed magnetic fields, in which strongly deformed resistance plateaus were observed \cite{Krstic08}. We show that the $\nu$ = 1 and 3 quantum Hall plateaus are missing in a magnetic field range up to $B$ = 53 T, which is consistent with the theoretical predictions \cite{Nomura06, Gusynin06}. We also show that room-temperature QHE can be observed at $B$ = 50 T even in a conventional-mobility device with $\mu$ = 4000 cm$^2$/Vs. The measurement technique established in this work can be used effectively for the study of half-integer QHE in single-layer graphene at high magnetic fields.

A single-layer graphene used in this study was extracted from Kish graphite and deposited onto a 300-nm-thick SiO$_2$ layer on top of a heavily doped Si wafer using the conventional mechanical exfoliation technique of graphite \cite{Novoselov04}. A sufficiently large single-layer graphene flake was selected using an optical microscope and multiple electrodes were fabricated using electron-beam lithography (Elionix ELS 7500), followed by the electron-beam evaporation of Au/Ti (40 nm/4 nm) [Fig. 1(a)]. The charge concentration $n$ of the graphene layer can be tuned by applying a gate bias voltage $V_\textrm{g}$ to the heavily doped silicon wafer, which serves as a global back gate. In order to remove surface impurities, the device was annealed in a measurement cryostat at $T$ $\sim$ 400 K in vacuum ($P$ $\sim$ 1 $\times$ 10$^{-2}$ Pa) for several hours prior to the measurements \cite{Novoselov05}. Figure 1(b) shows the resistivity $\rho$ of the graphene layer as a function of $V_\textrm{g}$. $\rho$ shows a sharp peak at $V_\textrm{g}$ = 5 V [Fig. 1(b)], indicating the Dirac neutrality point in this device $V_\textrm{dirac}$ = 5 V. The electron mobility of the graphene layer is $\mu$ = 4000 cm$^2$/Vs at $V_\textrm{g}$ = 32 V. 

The magnetic field was generated using a pulse magnet immersed in liquid nitrogen. The magnet was energized using a 1 MJ capacitor bank. The magnetic field was obtained by integrating the emf signal from a pick up coil [Fig. 1(c)]. During the pulses, the voltage signals $V_{xy}$ were recorded with DC sample currents of $i$ = $\pm$ 200 nA. The voltage signals are subtracted from each other to eliminate the field-induced background voltage, and $R_\textrm{H}$ is obtained by dividing the voltage signal by the current $i$ [Fig. 1(d)]. The trace of the $R_\textrm{H}$ vs $B$ is studied in a downward sweep because the change of the field is slower.

Figure 2(a) shows $R_\textrm{H}$ as a function of the inverse of the magnetic field $B^{-1}$ measured for an electron-injected device  ($V_\textrm{g}$ = 17.5 V) at $T$ = 4.2 K. In this measurement, the magnetic field was swept from $B$ = 53 T to 0 T. Hall plateaus are clearly visible at  $R_\textrm{H}$ = 13.3, 4.1, and 2.2 k$\Omega$ [dashed lines in Fig. 2 (a)], and these values correspond to the quantum Hall resistance of  $R_\textrm{H}$ = $h/2e^2$, $h/6e^2$, and $h/10e^2$, respectively, within 10\% accuracy. 
In general, the filling factor $\nu$ is proportional to carrier concentration and the inverse of the magnetic field $\nu$ = $n \phi _0 B ^{-1}$, where $\phi _0$ = 4.14 $\times$ 10$^{-15}$ Tm$^2$ is the flux quantum. The carrier concentration $n$ is calculated using the formula for graphene on a SiO$_2$ layer: $n$ = $\alpha$ ($V_\textrm{g}$ $-$ $V_\textrm{dirac}$), with a constant $\alpha$ = 7.2 $\times$ 10$^{10}$ cm$^{-2}$V$^{-1}$ determined from a theoretical estimation \cite{Novoselov05}. The filling factors at each Hall plateau center are calculated to be $\nu$ = 2.05, 5.96, and 10.01. These values are equal to the integer values of  $\nu$ = 2, 6, and 10, respectively, within experimental error. Therefore, the Hall plateaus observed in Fig. 2(a) are attributed to the QH plateaus with filling factors $\nu$ = $4(N + 1/2)$ where $N$ is an integer value. Here, we point out that this is the first experimental observation of half-integer QHE with reliable quantized Hall resistances in pulsed high magnetic fields. 

Figure 2(b) shows a plot of $R_\textrm{H}$ vs $B^{-1}$ measured for a hole-injected device ($V_\textrm{g}$ = $-$10 V) in sweeping magnetic fields from $B$ = 25 T to 0 T.  We observe the quantization of the Hall plateaus at $R_\textrm{H}$ = $(h/4e^2)(N + 1/2)^{-1}$ with $N$ = 1, 2, and 3 [Fig. 2(b)]. The filling factor at each Hall plateau center is calculated to be $\nu$ = $-2.1$, $-6.0$, and $-10.8$, respectively. Again, $R_\textrm{H}$ shows half-integer quantization, and this is the first experimental observation of half-integer hole QHE in hole-injected graphene in pulsed magnetic fields. 

Here, we emphasize the technical difficulties in measuring QHE under short ($\sim$ms) and strong pulsed magnetic fields. When a pulsed magnetic field is applied, an additional transient current is induced in the sample and measurement circuits. Therefore, even in a conventional semiconductor two-dimensional electron gas system, this transient current results in the deformation of quantum Hall plateaus \cite{Burgt92, Takamasu95}. To obtain well-developed Hall plateaus, Burgt \textit{et al.} theoretically compensated for the transient current flow in the device by assuming the appropriate capacitance\cite{Burgt92}. Kido \textit{et al.} used an active shielding technique; the voltages of the outer and inner conductors of the coaxial cables were equalized using the buffer amplifiers of unity gain \cite{Takamasu95, Kido82} to eliminate the capacitance of the circuits. In our study, we employed simple copper wires inside the measurement cryostat thus reducing the capacitance of the circuit as much as possible, and all the other electric cables and measurement equipment including a computer were installed in a shield box set near the magnet. This method has given sufficient low signal-to-noise ratio and suppresses the deformation effects.

Next, we study the $R_\textrm{H}$ vs $B$ up to 53 T for various carrier concentrations. Figure 3 shows the  $R_\textrm{H}$ vs $B $ measured at (a) $V_\textrm{g}$ = 17.5, (b) 30.0, (c) 40.0, and (d) 48.0 V. Under these conditions, the carrier concentrations are calculated to be (a) $n$ =  9.0$\times$10$^{11}$, (b) 1.8$\times$10$^{12}$, (c) 2.5$\times$10$^{12}$, and (d) 3.1$\times$10$^{12}$ cm$^{-2}$, respectively. Here, the plateaus in the $R_\textrm{H}$ are located exactly at the expected positions of the filling factors $\nu$ = 2, 6, and 10, as indicated by the red, green, and black arrows, respectively. This shows that the half-integer QHE is correctly measured for various gate bias voltages in our experiments.

Here, we discuss the $\nu$ = 3 QH plateau. In Fig. 3, the vertical dashed lines indicate the expected positions of the $\nu$ = 3 QH plateaus, and one can clearly see that the $\nu$ = 3 QH plateau is missing at any magnetic fields and any carrier concentrations studied. This result is in contrast to that of the previous study of the two-terminal magnetoresistance $R_\textrm{2p}$ using pulsed magnetic fields, in which Krsti\'c \textit{et al.} claimed that they have observed the $\nu$ = 3 QH state at $B$ = 30 $\sim$ 55 T at $n$ = 2.13 cm$^{-2}$ in a device with a lower mobility $\mu$ = 2,000 cm$^2$/Vs\cite{Krstic08}. In their study, the $\nu$ = 2 QH plateau, which is normally expected to be more robust than the $\nu$ = 3 QH plateau, was missing. On top of that, the interface resistance between a metal electrode and graphene, which is on the order of several kiloohms, could not be eliminated from $R_\textrm{2p}$, whereas the Hall resistance  in our measurements is correctly quantized to the expected quantum resistances within 10\% 
accuracy using a more reliable four-terminal geometry. Therefore, it is likely that the $\nu$ = 3 QH plateau claimed by Krsti\'c \textit{et al.} is attributed to the $\nu$ = 2 QH plateau associated with incorrectly quantized resistance. The absence of the $\nu$ = 3 QH plateau is consistent with the theoretical predictions;  Gusynin \textit{et al.} predicted that the $\nu$ = 3 QH state is absent at any magnetic fields \cite{Gusynin06}. Nomura \textit{et al.} estimated the critical values of $B$ for the $\nu$ = 3 QH plateau for the device with a mobility $\mu$ = 2000 cm$^2$/Vs to be $B$ = 1500 T \cite{Nomura06}, which is by far beyond the magnetic field reached in the present study.

In Fig. 3(a), the $\nu$ = 1 QH plateau is missing, although the magnetic field $B$ = 37.2 T corresponds to $\nu$ = 1 for $V_\textrm{g}$ = 17.5 V. By contrast, the $\nu$ = 1 QH plateau was observed in the high-mobility device with $\mu$ = 50,000 cm$^2$/Vs \cite{Zhang06} and $\mu$ = 20,000 cm$^2$/Vs \cite{Jiang07}. The lack of the $\nu$ = 1 QH plateau in this study is attributed to the relatively low mobility of the device, because $\mu \times B$ is less than the lower limit for observing the $\nu$ = 1 QH plateau \cite{Nomura06}.

We observe the plateau like structure in $R_\textrm{H}$ at $\nu$ = 4 ($B$ = 30 T) for $V_\textrm{g}$ = 48.0 V [Fig. 3(d)], which is reproducible for every set of pulses. Although $B$$\times$$\mu$ is smaller than the theoretically estimated lower limit for observing the $\nu$ = 4 QH state \cite{Nomura06} and the experimental observations of the $\nu$ = 4 QH state have been limited to ultrahigh-mobility devices in DC magnetic fields, this structure can be attributed to the $\nu$ = 4 QH state. Further studies of $R_{xx}$ and $R_\textrm{H}$ are required to clarify the plateau like structure. 

Finally, we conduct QHE measurements at elevated temperatures. Figures 4(a) and 4(b) show $R_\textrm{H}$ as a function of the $B$$^{-1}$ measured at (a) $T$ = 110 and (b) 290 K for (a) $V_\textrm{g}$ = 30 and (b) 40 V, respectively. Discernible QH plateaus at $\nu$ = 2 and 6 are observed [Fig. 4(a)] at $T$ = 110 K, and these plateaus become less pronounced but still survive up to room temperature ($T$ = 290 K)[Fig. 4(b)]. To date, room-temperature QHE has been observed only in a device with a high mobility $\mu$ $>$ 10,000 cm$^2$/Vs \cite{Novoselov07} at DC high magnetic fields. Our result shows that, using pulsed high magnetic fields, room-temperature QHE is observable even for a conventional-mobility device with $\mu$ = 4000 cm$^2$/Vs.

In summary, we have observed half-integer QHE in a graphene multiterminal device at magnetic fields of up to $B$ = 53 T using a pulse magnet. With either electron- or hole-type charge carriers, the Hall resistance $R_\textrm{H}$ is quantized into $R_\textrm{H}$ = $(h/e^2)\nu ^{-1}$ with the filling factors $\nu$ = $\pm$2, $\pm$6, and $\pm$10. Up to $B$ = 53 T, the $\nu$ = 1 and 3 QH plateaus were missing in our device, which can be attributed to the relatively low mobility of our device used in this study. At $B$ = 50 T, a $\nu$ = 2 QH plateau can be observed at room temperature ($T$ = 290 K) even with a conventional-mobility device $\mu$ = 4000 cm$^2$/Vs. We have shown that precise magnetotransport measurements of graphene can be conducted in an environment of pulse magnetic fields. This work is a step forward to investigating the electrical properties of graphene at much higher magnetic fields $B$ $>$ 100 T.

\begin{acknowledgments}
The authors thank Y. Hirayama, M. Kawamura, T. Osada, and P. Kim for helpful discussions and technical support. This work is supported by Grant-in-Aid from MEXT (No. 17244120), Special Coordination Funds for Promoting Science and Technology, and Grant-in-Aid for Scientific Research on Priority Areas, ``High Field Spin Science in 100T"(No.451) from MEXT. One of the authors (S. M.) acknowledges the JSPS Research Fellowship for Young Scientists. 
\end{acknowledgments}

\clearpage
\begin{figure}[t]
\begin{center}
\includegraphics[width=7cm]{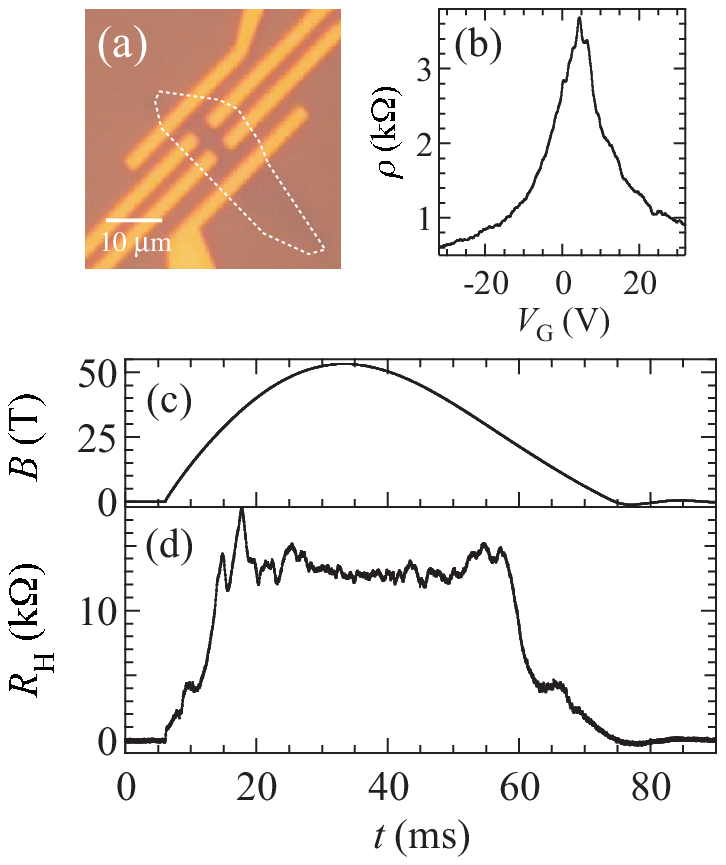}
\end{center}
\caption{(color online) (a) Optical image of the device. Outline of the graphene layer is highlighted by dashed lines. (b) Dependence of the resistivity $\rho$ on the gate bias voltage $V_\textrm{g}$ at $T$ = 4.2 K. (c) Time dependence of the magnetic field $B$. (d) Time dependence of the Hall resistance $R_\text{H}$ with $V_\textrm{g}$ = 30 V at $T$ = 4.2 K.}
\label{f1}
\end{figure}

\clearpage
\begin{figure}[t]
\begin{center}
\includegraphics[width=8cm]{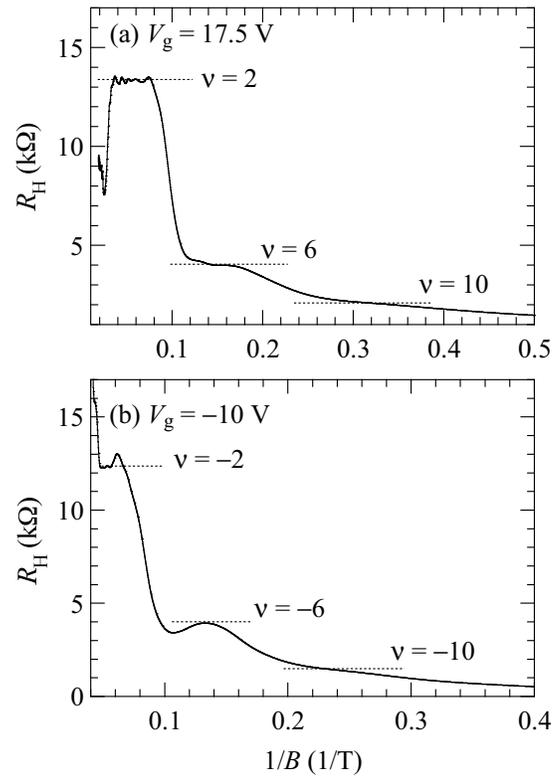}
\end{center}
\caption{Hall resistance $R_\textrm{H}$ as a function of the inverse of pulsed magnetic field $B^{-1}$ for gate bias voltages (a) $V_\textrm{g}$ = 17.5 V (electron doped) and (b) $-$10 V (hole doped) measured at $T$ = 4.2 K. The magnetic field was swept from $B$ = 53 to 0 T. Dotted lines indicate the $R_\textrm{H}$ at each Hall plateau.}
\label{f2}
\end{figure}

\clearpage
\begin{figure}[t]
\begin{center}
\includegraphics[width=8cm]{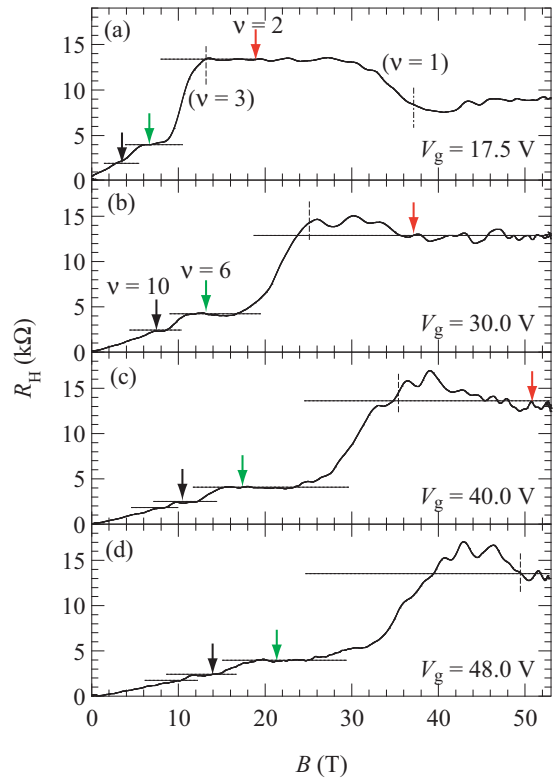}
\end{center}
\caption{(color online) Hall resistance $R_\textrm{H}$ as a function of magnetic field $B$ for gate bias voltages (a) $V_\textrm{g}$ = 17.5, (b) 30.0, (c) 40.0, and (d) 48.0 V at $T$ = 4.2 K. The horizontal dotted lines indicate the $R_\textrm{H}$ at each QH plateau. Arrows indicate the expected positions of the QH plateaus with $\nu$ = 2, 6, and 10. The vertical dashed (dashed and dotted) line indicates the expected position of the $\nu$ = 3 ($\nu$ = 1) QH plateau. }
\label{f3}
\end{figure}

\clearpage
\begin{figure}[t]
\begin{center}
\includegraphics[width=8cm]{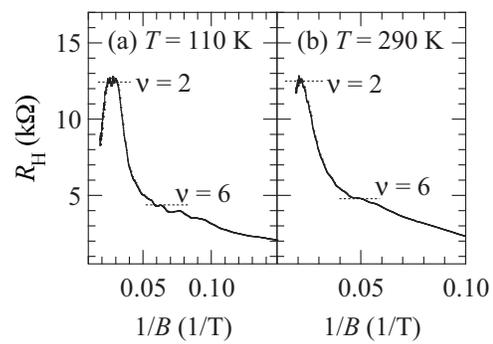}
\end{center}
\caption{Hall resistance as a function of the inverse of magnetic field $B^{-1}$ measured at  (a) $T$ =110 and (b) 290 K for gate bias voltages  (a) $V_\textrm{g}$ =30 and (b) 40 V.  The dotted lines indicate the $R_\textrm{H}$ at each Hall plateau.}
\label{f4}
\end{figure}

\end{document}